# Refractive Index Dispersion Fingerprinting via Scanning-Free Parallel Multi-Wavelength SPR on a Single Aluminum Film

Zihao Luo[a], Zhiying Chen[a], Yueqian Zhang[a], Xue Liu[b], Changsen Sun[a], Dmitry Kiesewetter[c], Sergey Krivosheev[c], Sergey Magazinov[c], Victor Malyugin[c], Xue Han[a,*]

[a] *School of Optoelectronic Engineering and Instrumentation Science, Dalian University of Technology, Dalian, 116024, China*

[b] *DUT-BSU Joint Institute, Dalian University of Technology, Dalian, 116024, China*

[c] *Institute of Energy, Higher School of High Voltage Engineering, Peter the Great St. Petersburg Polytechnic University, St. Petersburg, 195251, Russia*

[*] *xue_han@dlut.edu.cn*

**ABSTRACT:** Real-time characterization of refractive index (RI) dispersion is pivotal for advanced optical sensing, yet conventional surface plasmon resonance (SPR) platforms are bottlenecked by the narrow bandwidth of noble metals (Au, Ag) and the mechanical instability of sequential scanning. Here, we report a novel Al-based parallel SPR platform that overcomes the bandwidth and temporal constraints of conventional noble-metal systems. Leveraging the unique low-loss broadband response of Al, enabled by the suppression of interband transitions, we engineered a system for simultaneous excitation at 450, 520, and 635 nm. By integrating spectral-angle multiplexing with RGB-channel demultiplexing on a CMOS camera, the platform achieved acquisition of dispersion profiles without mechanical motion. Validated against NaCl solutions, the system demonstrates metrological accuracy and exceptional agreement with Cauchy dispersion models. The proposed architecture eliminates temporal drift and vibration errors, establishing a new benchmark for real-time dispersion characterization. By decoupling sensing from mechanical constraints, this work pioneers a compact, robust framework for next-generation, field-deployable sensors capable of distinguishing complex analytes via their unique spectral signatures.

# 1. INTRODUCTION

Accurate characterization of refractive index (RI) dispersion is fundamental to advances in material science [1, 2] and nano-photonics [3, 4]. However, existing techniques face a fundamental trade-off. Traditional methods are inherently limited to single wavelengths or static geometries [5, 6], while high-precision interferometry and ellipsometry preclude real-time monitoring of dynamic fluids due to complex phase retrieval and stability requirements [7-9]. Emerging fiber sensors, though sensitive, lack the broadband resolution for optical fingerprinting [8]. Surface plasmon resonance (SPR) is another physical phenomenon used for RI sensing. Recent SPR advances have pushed sensitivity limits but retain critical bottlenecks. Nanostructured metasurfaces enhance field confinement but require intricate nanofabrication and track only single resonance dips [10-13]. Moreau et al. [14] demonstrated a film-coupled silver nanocube array that achieves near-perfect absorption via gap-plasmon resonances without spatial patterning. But it requires complex colloidal synthesis and suffers from random assembly. Thus, current sensors are either spectrally limited (single-point, scanning) or structurally complex (nanopatterned, static).

To surmount these bottlenecks, we present a transformative, Al-based parallel SPR platform that redefines dynamic sensing through spectral-angle multiplexing. Unlike Au/Ag, Al supports SPR across the UV-NIR spectrum, enabling intrinsic broadband excitation. Our system employs a focused-beam angular interrogation to simultaneously excite three strategic wavelengths (450, 520, 635 nm). By RGB channel separation, we achieved single-shot, scanning-free acquisition of the RI dispersion profiles. This architecture uniquely circumvents the temporal drift of sequential scanning and eliminates the spectral bandwidth limits of noble metals. Consequently, this work establishes a robust framework for real-time dispersion characterization, setting a new possible benchmark for capturing transient changes in the RI dispersion in next-generation sensing.

# MATERIALS AND METHODS

### Theoretical Simulation

The exceptional performance of Al stems from a high bulk plasmon energy (14.98 eV) and a primary interband transition threshold at 1.544 eV (~800 nm) [15]. And the real part of dielectric function exhibits a smooth, monotonic negative variation across the visible spectrum. This ensures robust momentum matching for efficient SPR excitation over a continuous, ultra-broad bandwidth, enabling capabilities that are physically inaccessible to noble metals. Consequently, Al supports low-loss SPP propagation from the deep UV to the near-infrared (NIR), a spectral window where Au is severely compromised by a strong interband transition at 2.4 eV (~ 500 nm) [16, 17]. Nobel metal Ag has a strong interband transition at 4 eV (~ 300 nm) but suffers strong contamination due to the chemical activity. At regular room environment, the native oxide layer actually protect the Al metallic layer which is another advantage.

To rigorously optimize the sensor architecture for parallel, scan-free operation, we performed transfer matrix method (TMM) theoretical modelling of the Kretschmann configuration within classical local electrodynamics, a formalism valid for our planar-film geometry. As illustrated in Fig. 1 (a) – (c), Al films yield pronounced, deep SPR dips concurrently at 450, 520, and 635 nm over a well-defined thickness window, confirming feasible multi-wavelength SPR excitation. The native oxide layer was included in the multilayer stack to faithfully capture subwavelength interfacial optical effects. Unless otherwise specified, all quoted thicknesses refer exclusively to the metallic Al layer. In sharp contrast, Au films (Fig. 1(d)-(f)) fully suppress SPR excitation at 450 nm and 520 nm. This originates from intrinsic physical limitations as mentioned previously. Though Au sustains SPR at 635 nm (Fig. 1(f)), its absence of blue-green spectral response renders it incompatible with our parallel acquisition scheme. For simulations, complex RIs of Al were acquired from ellipsometry measurements to ensure fidelity, while optical constants of Au were adopted from the well-known Johnson and Christy dataset [17]. **Supporting Information (SI) Section 1** further corroborates unique suitability of Al for spectral-angle multiplexing via comprehensive comparisons of broadband resonance profiles at fixed incident angles. TMM simulation accounted explicitly for the wavelength-dependent dispersion relations of the fused silica substrate and the native alumina layer.

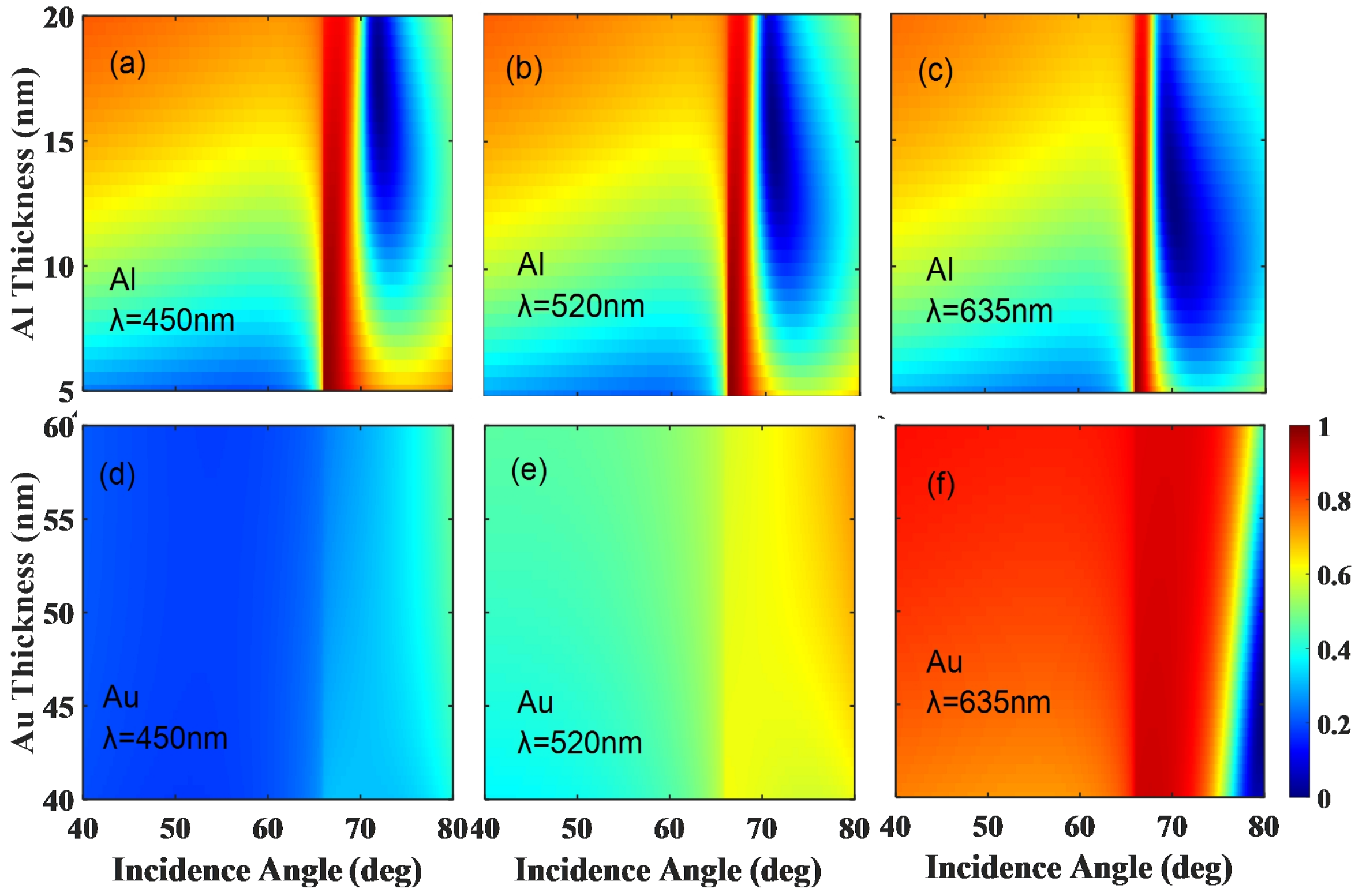

Figure 1: TMM calculated angular reflectivity for varied thicknesses of Au and Al films. Al at (a) 450, (b) 520 and (c) 635 nm excitation. Gold at (d) 450, (e) 520 and (f) 635 nm excitation.

Building upon the broadband resonance characteristics identified in Fig. 1, an optimization of the incident angular range and Al film thickness was conducted to maximize the multi-wavelength performance. Specifically, Al films with thicknesses of 16, 17, and 18 nm were theoretically analyzed across a range of surrounding medium RI to evaluate their angular sensitivity (detailed in **SI Section 2, Fig. S2**). The incident angular range was defined to span from 66° to 75°, ensuring coverage of the resonance dips for all three target wavelengths. Theoretical angular sensitivities for these three thicknesses were systematically compared (**SI Section 2, Fig. S3**). Through this comprehensive analysis, 17 nm was identified as the optimal thickness of the Al layer. While a marginal reduction in the sensitivity at 450 nm resulted (126.68°/RIU), the peak theoretical sensitivity for the 520 nm (121.69°/RIU) and 635 nm (116.09°/RIU) channels were yielded.

**Deposition of Aluminum Film**

High-quality Al films were fabricated on fused silica (quartz) substrates via high-vacuum resistive evaporation using a TECHNOL ZHD300 system. To ensure the formation of high-purity metallic films with minimal contamination, the deposition was performed under a base pressure of $5 \times 10^{-4}$ Pa with a rapid deposition rate [18]. The precise thickness and complex RIs of both the metallic Al layer and its native alumina capping layer were characterized using ellipsometry (J.A. Woollam, M-2000V). The detailed optical constants and fitting parameters are provided in **Supporting Information (SI) Section 3**. The ellipsometric analysis confirmed the deposition of a 17.16 nm metallic Al layer capped by a naturally formed 3 nm oxide layer. The native oxide layer was stable in the aqueous environment for days and in a regular room environment for months without further growth of the oxide layer observed.

**Preparation of NaCl Solutions and SPR Sensing Cell**

Aqueous sodium chloride (NaCl) solutions were prepared at weight percentages (wt%) of 0%, 0.3%, 0.9%, and 1.5% to serve as analytes with varying RIs. For comparative validation, the bulk RI of each solution was measured at room temperature using a standard Abbe refractometer (INESA, WYA-2S) equipped with a polychromatic white light source, resulting an effective RI.

The SPR sensing cell was constructed by bonding the Al-coated fused silica substrate to a glass slide using double-sided adhesive tapes, thereby forming a sealed, microfluidic flow-through cell. Two integrated inlet and outlet ports were incorporated into the cell design to enable continuous, bubble-free solution exchange.

**Parallel Multi-wavelength SPR Setup**

The experimental realization of the parallel multi-wavelength sensing platform was based on the Kretschmann configuration, as schematically illustrated in Fig. 2. The optical architecture integrates three semiconductor laser sources operating at 450, 520, and 635 nm. These distinct wavelengths are co-linearly combined into a single beam path via a series of dichroic mirrors, enabling simultaneous excitation without temporal multiplexing. The composite beam is subsequently purified to pure transverse magnetic (TM) polarization using a polarizing prism, a critical requirement for efficient SPR excitation.

To facilitate angular multiplexing, the collimated beams were expanded to diameter of 16 mm using a 5 × beam expander (LBTEK, FBE-5X-A). These expanded beams were then focused onto the base of a fused silica semi-cylindrical prism by an achromatic lens ($f$ = 75 mm). And the focal point was adjusted to the interface between the Al and surrounding medium. This optical arrangement maps the spatial profile of the incident beam to a continuous angular spectrum at the metal-dielectric interface, spanning a range of approximately 12.2°. This range satisfied the 9° span of incident angle requirement (lower 66° to upper 75°). The scanning-free angular interrogation was achieved for these three wavelengths.

The sensing cell was mounted on a precision rotation stage (Zolix, RauK20-100, resolution 0.0005°). The sensing interface between Al and the surrounding medium was alignment with the central axis of the rotation stage. This procedure guarantees the angle-pixel transfer accuracy. The stage was held strictly stationary during SPR sensing.

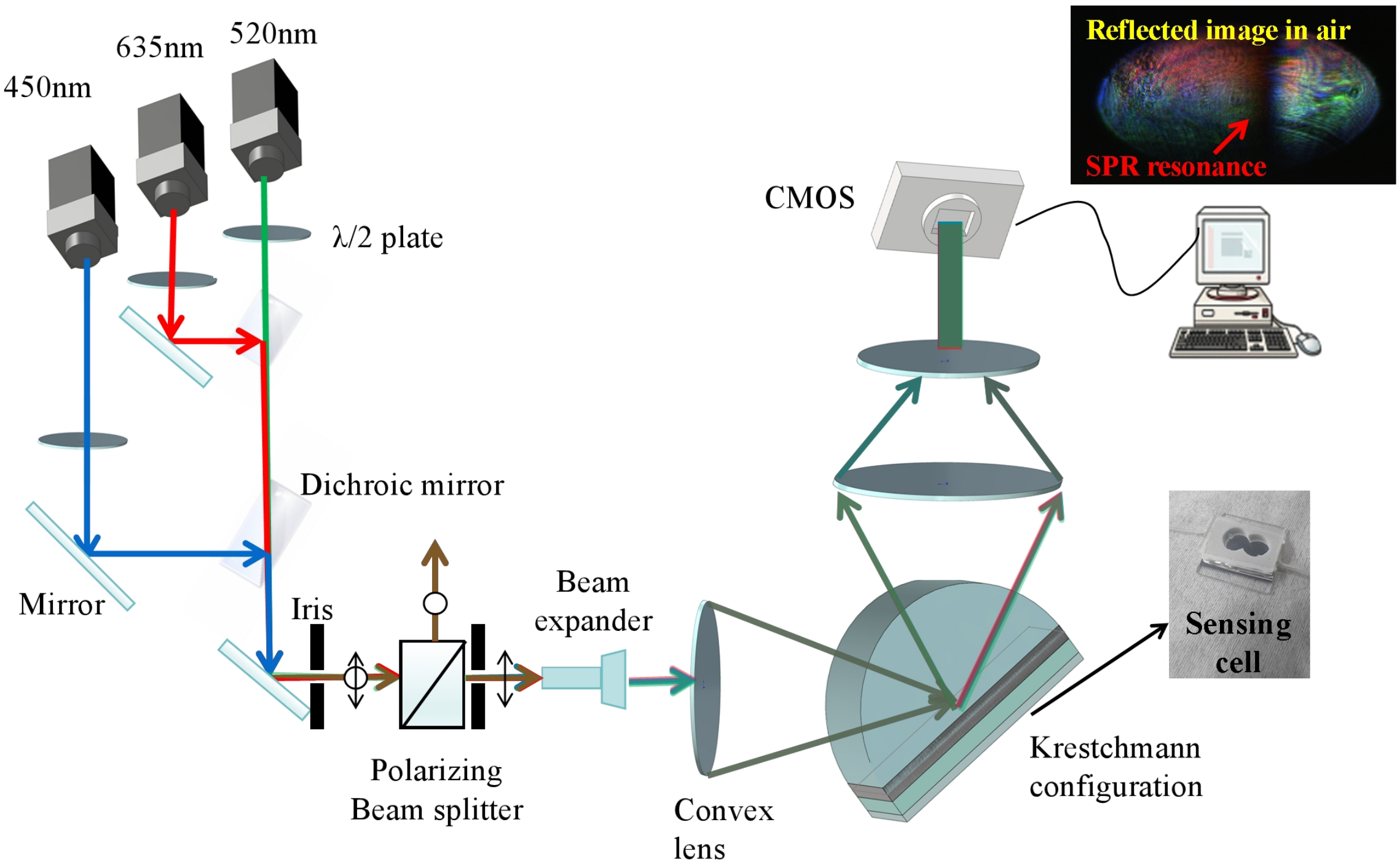

Figure 2: Schematic diagram of experimental setup

The reflected beams were collimated and relayed onto a CMOS camera (Thorlabs DCC3240C, active area $6.78\ \text{mm} \times 5.43\ \text{mm}$) through a dual-lens system ($f_1 = 75$ mm, $f_2 = 100$ mm). All optical components along the detection path were achromatic, suppressing chromatic aberration. The spatially distributed reflected intensity captured by the camera was converted to incident-angle spectra using a calibrated pixel-to-angle mapping routine. System angular resolution, defined as the angular increment per pixel, was quantitatively evaluated for each wavelength channel (**SI Section 4**), yielding 0.0068°/pixel at 450 nm, 0.0073°/pixel at 520 nm, and 0.0063°/pixel at 635 nm. To verify signal fidelity and guarantee independent readout among parallel detection channels, we further performed systematic characterizations of RGB crosstalk ratio (<12%, **SI Section 5**).

**Measurement Procedure**

The absolute incident angle calibration of the scanning-free interrogation system was established by aligning the reflected beam with the incident optical path, defining this zero-reflection geometry as the 0° reference for the rotation stage. The SPR response was initially validated in air by rotating the stage to 45° based on the RIs of fused silica and air. A distinct resonance dip was clearly observed (top-right inset, Fig. 2), confirming the successful excitation of SPPs on the Al film.

Following validation, deionized (DI) water was introduced into the sensing cell as the baseline analyte. The rotation stage was then precisely adjusted to a position the resonance dip of DI water within the low-angle region of the field of view. This strategic angular positioning was critical. Since DI water possesses the lowest RI among all tested samples, its resonance angle represents the lower bound of the sensing range. Consequently, resonance dips for all subsequent NaCl solutions with higher RIs would naturally shift toward larger angles, ensuring that the entire dynamic range of the sensor remains fully contained within the capture window without truncation. And the incident intensity of each laser beam was individually optimized to maximize the signal-to-noise ratio while avoiding camera. With the optimal central incident angle and beam intensities fixed, NaCl solutions of increasing concentration were injected sequentially into the sensing cell. The dynamic evolution of the resonance was captured in real-time via video acquisition and the exposure time was 9 μs for each frame.

**DATA ANALYSIS**

The raw images captured by the CMOS sensor were spectrally decomposed into RGB channels to isolate the signals corresponding to the 635, 520, and 450 nm wavelengths, respectively. To ensure strict spatial consistency across the parallel detection channels, identical pixel rows (260-370) were selected for analysis in all three channels. The reflected intensity profile for each wavelength was then derived by averaging the pixel values within these defined rows, effectively suppressing random pixel-to-pixel noise (Fig. 3b-d). To calculate the normalized reflectivity ($R = \frac{I}{I_0}$), the maximum intensity ($I_0$) observed in each frame was utilized as the reference baseline. For presentation, the intensity profiles were smoothed using a Savitzky-Golay filter. This filter was not used for data analysis.

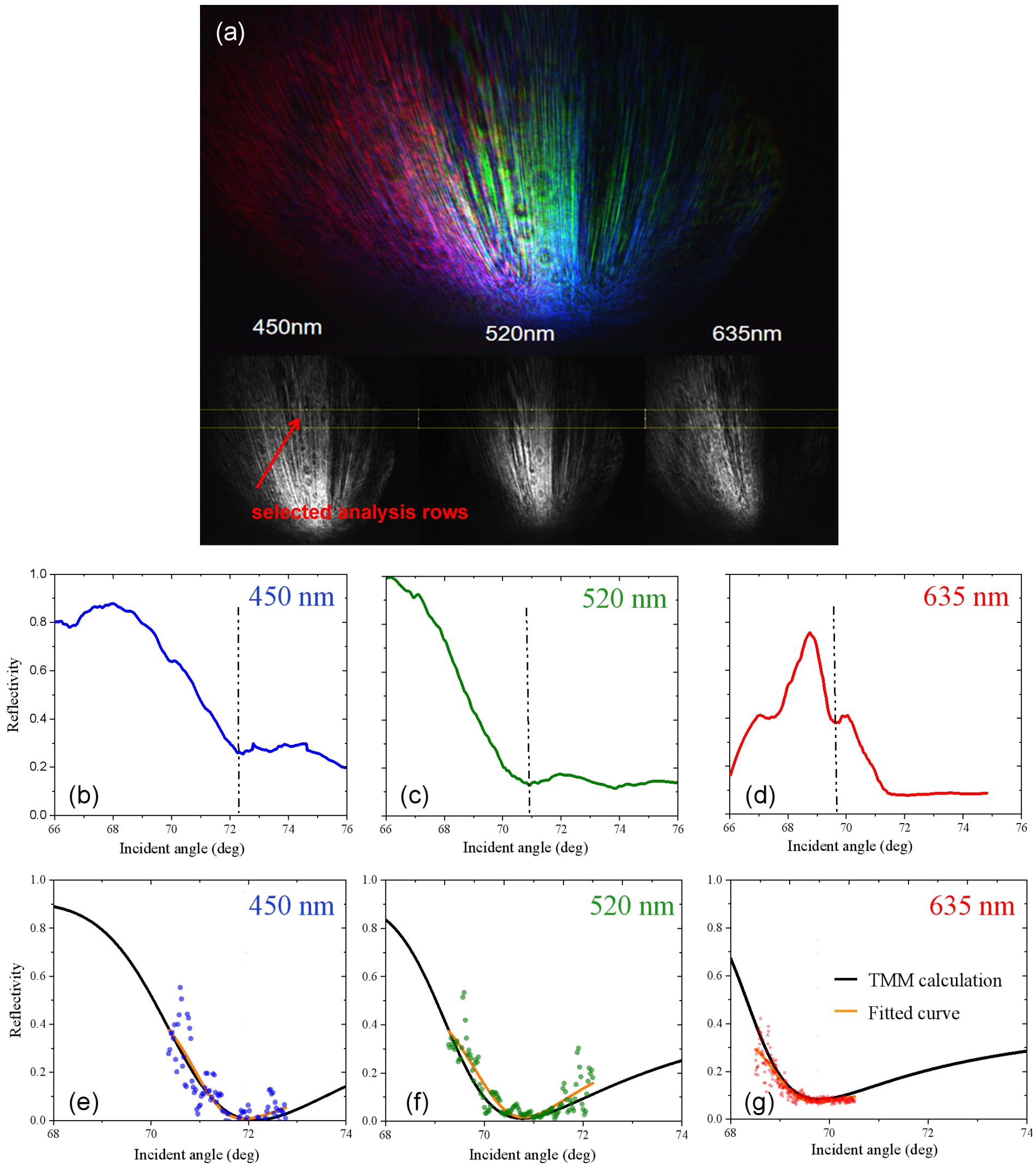

Figure 3: (a) Reflected image for 0.9% NaCl solution. Three RGB channels are presented separately for (b) 450, (c) 520, and (d) 635 nm. The experimental (scatter), fitted SPR curve (orange line), and TMM calculated curves (black lines) are presented for (e) 450, (f) 520, and (g) 635 nm.

The experimental SPR curve was first fitted with Eqs. (1) and (2), capturing the resonance angle. In this model, $R$ is the reflectivity, $A$ is the peak intensity, $B$ and $C$ govern the curve's symmetry and asymmetry, $D$ denotes the resonance angle, and $E$ characterizes the dip width [19]. The known fitting parameters include the surface plasmon wavevector ($K_x$), the substrate dielectric constant ($\varepsilon_{sub}$), the vacuum wavelength ($\lambda$), and the incident angle ($\theta_{in}$). Excellent agreement between the experimental data and the fitted curves is demonstrated in Fig. 3(e)-(g).

$$R = A(1 - \frac{B + C(K_x - D)}{(K_x - D)^2 + E^2}) \quad (1)$$

$$K_x = \sqrt{\varepsilon_{sub}} \frac{2\pi}{\lambda} \sin\theta_{in} \quad (2)$$

$$n_s = A' + \frac{B'}{\lambda^2} \quad (3)$$

The experimentally extracted resonance angles were integrated with the TMM model to retrieve the RI values of the NaCl solutions. This inversion process was executed via a custom MATLAB fitting algorithm that minimizes the deviation between the measured angular shifts and the theoretical resonance conditions. The resulting RI retrieval curves are presented in Fig. 3(e)-(g), which demonstrate an exceptional agreement between three distinct data sources: the raw experimental measurements, the analytical curve fitting to retrieve resonance angles, and the TMM simulation to retrieve RI values of the surrounding media.

# RESULTS & DISCUSSION

Comprehensive fittings for all tested wt % NaCl solutions are provided in **SI Section 6**. The retrieved RI values for NaCl solutions, plotted in Fig. 4(a), exhibit a robust linear dependence on the solute wt %, confirming the fidelity accuracy of the system. Consistent with the fundamental principles of normal dispersion, the measured RI demonstrates a distinct wavelength-dependency, with values monotonically decreasing as wavelength increases ($n_{450} > n_{520} > n_{635}$). To validate the reliability of these measurements, the retrieved RIs were cross-referenced with readings from a standard Abbe refractometer. As expected, this reference value from Abbe instrument (white light source) falls between the experimentally determined values for the 520 nm and 635 nm channels. This alignment confirms that the multi-wavelength system provides the dispersion relationship that is inherently averaged out in conventional white-light measurements.

We referenced the comprehensive study by Li et al., which reported the optical properties of NaCl solutions across the 400-700 nm . Their work established a benchmark RI increment of $1.8 \times 10^{-4}$ to $2.1 \times 10^{-4}$ per wt % increase in NaCl concentration [20]. Our experimental measurements yielded an RI increment range of $1.7 \times 10^{-4}$ to $2.1 \times 10^{-4}$ per wt % increase, demonstrating exceptional quantitative agreement with the established literature benchmarks. Crucially, by focusing on the concentration-dependent RI gradients ($\Delta n/\Delta wt\%$) rather than absolute RI values, the influence of the baseline solvent RI (DI water) is inherently eliminated.

Figure 4(b) illustrates the linear relationship between the resonance angle shift ($\Delta\theta_{res}$) and the RI variation ( $\Delta n_{sol}$ ) for the three excitation wavelengths. Linear regression analysis yields experimental angular sensitivities ( $S_{\theta} = \Delta\theta_{res}/\Delta n_{sol}$ ) that are remarkably consistent across the visible spectrum. This spectral uniformity underscores the unique broadband plasmonic response of the single Al film. Furthermore, the achieved sensitivity of the Al-based sensor is directly competitive with state-of-the-art Au-based SPR chips (other wavelength range) employing angular interrogation [18]. This result is particularly significant as it demonstrates that Al can match the performance of Au in the visible range while simultaneously enabling multi-wavelength operation, a capability that is physically precluded for Au in the blue-green spectrum.

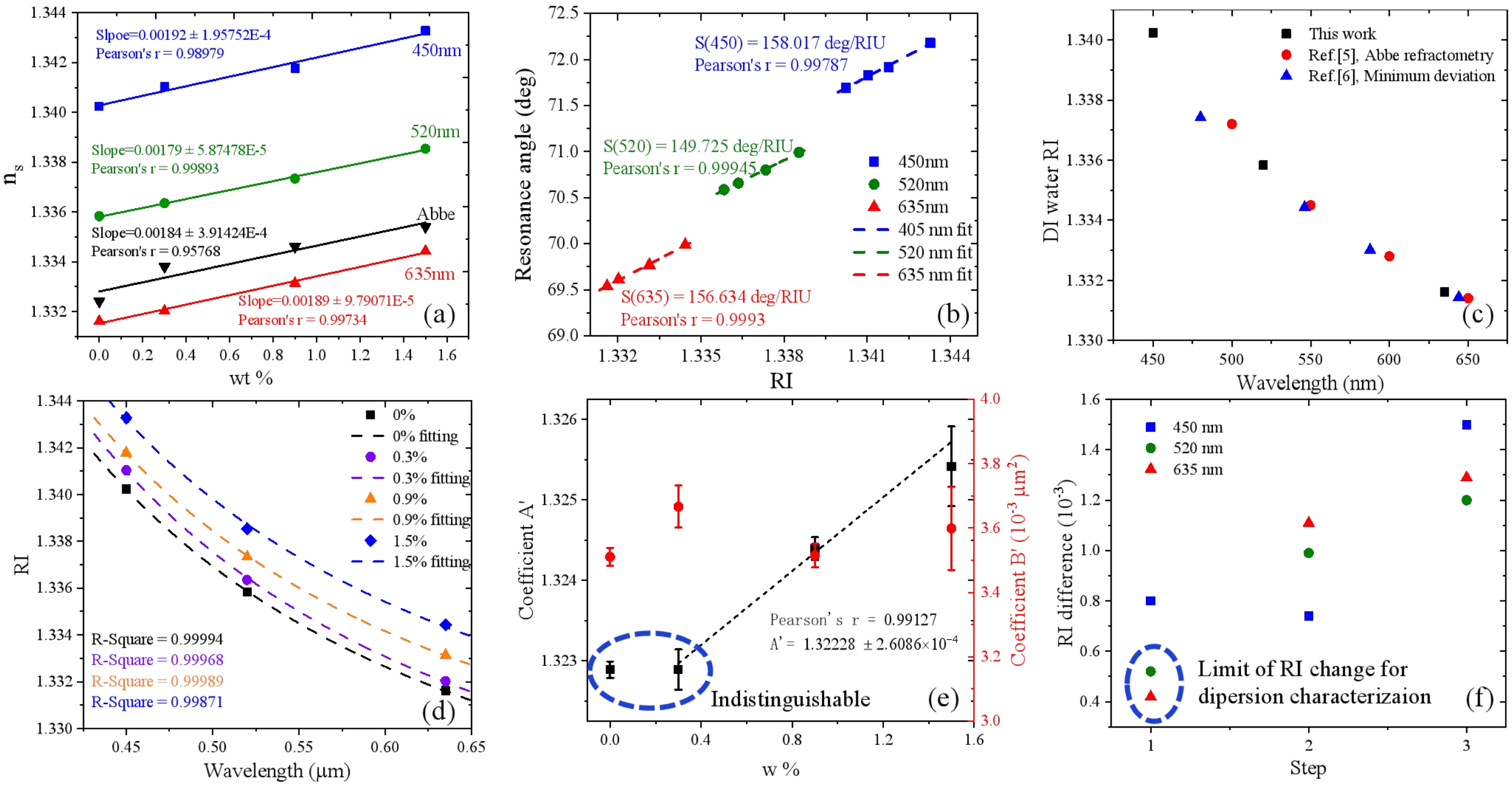

Figure 4. (a) Retrieved RI versus NaCl wt %. (b) Resonance angle shift versus retrieved RI for the blue, green, and red channels. (c) RI of DI water comparison among different research works. (d) Cauchy dispersion fitting for NaCl solutions at varied wt %. (e) Cauchy coefficients $A'$ and $B'$ versus NaCl wt%. (f) Minimum detectable RI change

The RI of DI water from our work matches other research works well, as shown in Fig. 4(c). The experimentally retrieved RI values at λ = 450, 520, and 635 nm showed exceptional agreement with established reference data [5, 6]. The experimentally determined RIs were successfully modeled using the Cauchy dispersion equation (Eq. 3) across all NaCl wt%, yielding high-quality fits with minimal residuals (Fig. 4(d)). Fig. 4(e) illustrates the evolution of the fitted Cauchy coefficients, $A'$ and $B'$, as a function of solute concentration, with error bars derived from the fitting uncertainty. While coefficients $A'$ for the 0% and 0.3% solutions are statistically indistinguishable within the error margins, a robust linear dependence is clearly evident for concentrations ranging from 0.3% to 1.5%. This linear trend confirms that $A'$ serves as a direct, quantitative proxy for the solute content in this regime. Coefficient $B'$, which governs the strength of the wavelength-dependent dispersion ($dn/d\lambda$) showing no statistically significant trend.

Figure 4(f) illustrates the limit of the system in dispersion characterization. The minimum resolvable RI difference between adjacent NaCl concentrations (0% vs. 0.3%) is approximately $10^{-4}$ RIU. With limited wavelength data points (three wavelengths), the fitting of the Cauchy equation cannot be extended to higher terms, which increases the error in $A'$ and $B'$. This is also the reason that coefficient $B'$ is almost a constant among different NaCl wt % solutions. This limitation can be improved by altering the detection components into physically separated channels using gratings or a hyperspectral camera.

## CONCLUSION

This study establishes a transformative, single Al-based parallel multi-wavelength angular interrogation SPR platform that achieves scanning-free, real-time RI dispersion fingerprinting across the visible spectrum. By synergizing multi-wavelength excitation with spectral-angle multiplexing via RGB-channel demultiplexing, the system simultaneously retrieves dispersion profiles. This architecture fundamentally circumvents the temporal variations and mechanical instability, inherent to conventional sequential SPR systems, enabling high-fidelity monitoring of dynamic fluidic events in real time. Validated against NaCl standards, the platform demonstrates metrological accuracy while leveraging the unique broadband, low-loss response of Al to overcome the spectral constraints of noble metals.

The proposed architecture is inherently modular and scalable. While currently operating in the visible regime, the core parallel interrogation strategy is universally applicable. Future iterations can expand spectral coverage into the UV or NIR by optimizing film thickness via TMM simulations. Upgrading the detection module to hyperspectral imaging will further elevate the system from discrete multi-wavelength sensing to continuous high-resolution spectral mapping, unlocking advanced capabilities for complex material identification. Regarding the exposure time, real-time measurement can be achieved by using much faster detectors. Currently, the incident intensity was limited by the saturation condition of the camera.

Looking forward, strategic enhancements will target single-molecule sensitivity and intelligent deployment. Refining angular resolution through spatial mode filtering (e.g.,pinhole integration) will sharpen resonance features, enabling the detection of subtle bio-interface dispersion changes (e.g., protein conformational shifts). Coupled with edge-computing pipelines for instantaneous data reconstruction, this platform will transition from a laboratory instrument to a field-deployable, intelligent sensing engine. This work bridges the gap between high-fidelity spectroscopic characterization and real-time, portable sensing. By decoupling spectral resolution from mechanical scanning, the proposed platform establishes a robust framework for next-generation optical fingerprinting, enabling applications in point-of-care diagnostics and in-situ environmental monitoring.

**Credit authorship contribution statement**

**Zihao Luo:** Experiments conduction, Data analysis, Writing-original draft.

**Zhiying Chen:** TMM calculation, SPR curve calibration.

**Yueqian Zhang:** Aluminum deposition and thickness calibration.

**Xue Liu:** Visualization.

**Changsen Sun:** Designed the protocol, Writing-review & editing

**Dmitry Kiesewetter:** Validation, Writing-review & editing

**Sergey Krivosheev:** Validation, Writing-review & editing

**Sergey Magazinov:** Code writing, Writing-review & editing

**Victor Malyuginc:** Validation, Writing-review & editing

**Xue Han:** Methodology, Supervision, Funding acquisition, Writing-review & editing.

**Declaration of Competing Interest**

The authors declare that they have no known competing financial interests or personal relationships that could have appeared to influence the work reported in this paper.

## Data availability

Data will be made available on request.

## Acknowledgments

The authors acknowledge the financial support from the National Key Research and Development Program of China (2024YFE0213500), the National Natural Science Foundation of China (Grant Nos. 62105053), Instrumental Analysis Center at Dalian University of Technology.

## References

1. Zhang, X.N., et al., *Complex refractive indices measurements of polymers in visible and near-infrared bands.* Applied Optics, 2020. **59**(8): p. 2337-2344.
2. Munkhbat, B., et al., *Optical Constants of Several Multilayer Transition Metal Dichalcogenides Measured by Spectroscopic Ellipsometry in the 300–1700 nm Range: High Index, Anisotropy, and Hyperbolicity.* ACS Photonics, 2022.
3. Perkins, J., et al., *Color Tunable, Lithography-Free Refractory Metal-Oxide Metacoatings with a Graded Refractive Index Profile.* Nano Lett, 2023. **23**(7): p. 2601-2606.
4. Chen, W.T., et al., *Dispersion-engineered metasurfaces reaching broadband 90% relative diffraction efficiency.* Nat Commun, 2023. **14**(1): p. 2544.
5. Kedenburg, S., et al., *Linear refractive index and absorption measurements of nonlinear optical liquids in the visible and near-infrared spectral region.* Optical Materials Express, 2012. **2**(11): p. 1588-1611.
6. Daimon, M. and A. Masumura, *Measurement of the refractive index of distilled water from the near-infrared region to the ultraviolet region.* Applied Optics, 2007. **46**(18): p. 3811.
7. Wang, Q., et al., *Optimization of cascaded fiber tapered Mach–Zehnder interferometer and refractive index sensing technology.* Sensors and Actuators B: Chemical, 2016. **222**: p. 159-165.
8. Yang, X., et al., *Simultaneous measurement of liquid level and refractive index based on a sandwich multimode optical fiber structure.* Optics & Laser Technology, 2024. **168**: p. 109856.
9. Zhu, P., et al., *Measurement of Refractive Index and Thickness of Thin Films Via Polarization-Projection Interferometry.* Laser & Photonics Reviews, 2025. **20**(3).
10. Wang, L., et al., *Refractive index and thickness measurements with ultrahigh sensitivity via versatile surface plasmon resonance holographic microscope.* Light: Advanced Manufacturing, 2026. **7**(0): p. 1.
11. Lodewijks, K., et al., *Boosting the figure-of-merit of LSPR-based refractive index sensing by phase-sensitive measurements.* Nano Letters, 2012. **12**(3): p. 1655-9.
12. Kramadhati, S., Y.C. Choi, and C.R. Kagan, *Large-Area, Narrow-Gap Plasmonic Nanodimer Metasurfaces Exploiting Colloidal Nanocrystals: Promising Platforms for Refractive Index Sensing.* ACS Applied Nano Materials, 2025.
13. Lin, L. and Y. Zheng, *Engineering of parallel plasmonic-photonic interactions for on-chip refractive index sensors.* Nanoscale, 2015. **7**(28): p. 12205-14.

14. Moreau, A., et al., *Controlled-reflectance surfaces with film-coupled colloidal nanoantennas.* Nature, 2012. **492**(7427): p. 86-+.
15. Rakic, A.D., et al., *Optical properties of metallic films for vertical-cavity optoelectronic devices.* Applied Optics, 1998. **37**(22): p. 5271-5283.
16. Olmon, R.L., et al., *Optical dielectric function of gold.* Physical Review B, 2012. **86**(23).
17. Johnson, P.B. and R.W. Christy, *Optical Constants of the Noble Metals.* Physical Review B, 1972. **6**(12): p. 4370-4379.
18. He, C., et al., *Sensitive Aluminum SPR Sensors Prepared by Thermal Evaporation Deposition.* ACS Omega, 2023. **8**(45): p. 43188-43196.
19. Kurihara, K., K. Nakamura, and K. Suzuki, *Asymmetric SPR sensor response curve-fitting equation for the accurate determination of SPR resonance angle.* Sensors and Actuators B-Chemical, 2002. **86**(1): p. 49-57.
20. Li, X., et al., *Optical Properties of Sodium Chloride Solution Within the Spectral Range from 300 to 2500 nm at Room Temperature.* Applied Spectroscopy, 2015. **69**(5): p. 635-40.